\documentclass[reprint,twocolumn,final,aps,prb,showpacs,superscriptaddress,amsmath,amssymb,amsfonts,floatfix]{revtex4-2}

\usepackage{graphicx}
\usepackage{amsmath}
\usepackage{listings}
\usepackage{xcolor}
\usepackage{bm}
\usepackage{ulem}
\usepackage{color}
\usepackage{mathrsfs}
\usepackage{textcomp, gensymb}
\usepackage{upgreek}
\usepackage[colorlinks=true,linkcolor=blue,citecolor=blue]{hyperref}

\usepackage{stackrel}
\usepackage{floatrow}
\usepackage{subcaption}

\usepackage{multirow}
\usepackage{url}

\usepackage{tabularx}
\usepackage{sidecap}
\usepackage{soul}
\usepackage{float}

\usepackage{orcidlink}
\begin{document}

\title{ChatGPT-Assisted Visualization of Atomic Orbitals: Understanding Symmetry, Mixed State, and Superposition}

\author{Huiping Han}
\affiliation{Faculty of Materials Science and Engineering, Kunming University of Science and Technology, Kunming, 650093, Yunnan, China}
	
\author{Liang Wu\orcidlink{0000-0003-1030-6997}}
\email{liangwu@kust.edu.cn}
\affiliation{Faculty of Materials Science and Engineering, Kunming University of Science and Technology, Kunming, 650093, Yunnan, China}

\date{\today}

\begin{abstract}
For undergraduate students newly introduced to quantum mechanics, solving simple Schr\"{o}dinger equations is relatively straightforward. However, the more profound challenge lies in comprehending the underlying physical principles embedded in the solutions. During my academic experience, a recurring conceptual difficulty was understanding why only $s$ orbitals, and not others like $p$ orbitals, exhibit spherical symmetry. At first glance, this seems paradoxical, given that the potential energy function itself is spherically symmetric. Specifically, why do $p$ orbitals adopt a dumbbell shape instead of a spherical one? For a hydrogen atom with an electron in the $2p$ state, which specific $2p$ orbital does the electron occupy, and how do the $x$, $y$, and $z$ axes in $2p_x$, $2p_y$, and $2p_z$ connect to the real world? Additionally, is the atom still spherically symmetric in such a state? These questions relate to core concepts of quantum mechanics concerning symmetry, mixed state, and superposition. This paper delves into these questions by investigating this specific case, utilizing the advanced visualization capabilities offered by ChatGPT. This paper underscores the importance of emerging AI tools in enhancing students' understanding of abstract principles.
\end{abstract}

\maketitle
\section{Introduction}
Quantum mechanics is a cornerstone of modern physics, offering profound insights into the nature of matter and energy at the smallest scales. However, for undergraduate students who are new to this area of study, the journey from solving basic Schr\"{o}dinger equations to fully grasping the physical meaning behind those solutions is often fraught with challenges. Although the mathematical mechanics of these equations can be mastered with practice, the conceptual leap required to understand the physical implications, particularly those related to orbital shape, symmetry, and superposition, can be much more elusive.

A recurring challenge faced by students pertains to understanding why only $s$ orbitals exhibit spherical symmetry, while $p$ orbitals adopt distinctly non-spherical, dumbbell shapes. 
This seems counterintuitive at first glance, especially given that the potential energy function in these systems is itself spherically symmetric.

These challenges are not merely academic; they strike at the heart of how students conceptualize the quantum world. For instance, in the case of a hydrogen atom with an electron in the $2p$ state, it is crucial to understand why the electron's probability distribution no longer retains spherical symmetry. This raises further questions: which specific $2p$ orbital does the electron populate, and how do the orientations of $2p_x$, $2p_y$, and $2p_z$ orbitals relate to the real space defined by the $x$, $y$, and $z$ axes? Additionally, students often wonder if the atom as a whole retains any form of spherical symmetry when in such states, or if this symmetry is completely broken.
This collection of questions delves into  essential quantum mechanics concepts, including symmetry, mixed states, and the superposition principle. 
They also highlight the importance of distinguishing between the mathematical representation of quantum states and their physical interpretations. Understanding these principles is crucial for students, as it forms the foundation upon which more advanced topics in quantum mechanics are built.

In this paper, by utilizing the advanced visualization capabilities offered by recent emerging AI tools, ChatGPT, we address these conceptual challenges by exploring the specific case of $p$ orbitals in the hydrogen atom, providing a detailed examination of the symmetry properties of these orbitals and their implications. The role of AI tools in physics education has already captured significant attention of the physics community \cite{PhysRevPhysEducRes.19.020003}.
The ability of AI tools to dynamically illustrate complex ideas offers a promising avenue for enhancing students' comprehension of difficult concepts in quantum mechanics. As such, this exploration not only addresses specific academic challenges, but also underscores the broader potential of emerging technologies to transform the way abstract scientific principles are taught and understood.

\section{ChatGPT-Assisted Visualization of Atomic Orbitals}

\subsection{Solving $2p$ orbitals for hydrogen atom}

We first request ChatGPT4o to solve the $2p$ orbitals, or we can find the solutions in any quantum chemistry textbook, e.g. Ref. \cite{levine2014}.

\begin{quote}
\textbf{User:} Please solve the $2p$ orbitals for the hydrogen atom. \\
\textbf{ChatGPT:} To solve the $2p$ orbitals for the hydrogen atom, we need to solve the Schrödinger equation for the hydrogen atom and specifically look at the solution for the $2p$ orbitals. The general form of the time-independent Schrödinger equation in spherical coordinates is:
\[
\hat{H} \psi(r, \theta, \phi) = E \psi(r, \theta, \phi)
\]
Where:
\begin{itemize}
    \item $\hat{H}$ is the Hamiltonian operator for the hydrogen atom, which includes kinetic energy and Coulomb potential terms.
    \item $\psi(r, \theta, \phi)$ is the wavefunction in spherical coordinates.
    \item $E$ is the energy eigenvalue corresponding to a particular quantum state.
\end{itemize}

\textbf{Step 1: Quantum Numbers}

For the hydrogen atom, the wavefunction $\psi(r, \theta, \phi)$ is characterized by three quantum numbers:
\begin{itemize}
    \item \textbf{Principal quantum number} ($n$): For the $2p$ orbitals, $n = 2$.
    \item \textbf{Azimuthal (orbital angular momentum) quantum number} ($l$): For $p$ orbitals, $l = 1$.
    \item \textbf{Magnetic quantum number} ($m$): For p orbitals, $m = -1, 0, +1$.
\end{itemize}

\textbf{Step 2: Separation of Variables}

The Schrödinger equation in spherical coordinates is separable into radial and angular parts:
\[
\psi_{n,l,m}(r, \theta, \phi) = R_{n,l}(r) Y_{l,m}(\theta, \phi)
\]
Where:
\begin{itemize}
    \item $R_{n,l}(r)$ is the radial part of the wavefunction.
    \item $Y_{l,m}(\theta, \phi)$ are the spherical harmonics, which represent the angular part.
\end{itemize}

\textbf{Step 3: Radial Part $R_{n,l}(r)$}

For the $2p$ orbital, the radial part $R_{2,1}(r)$ is given by:
\[
R_{2,1}(r) = \frac{1}{2\sqrt{6a_0^3}} \left(\frac{r}{a_0}\right) e^{-r/2a_0}
\]
Where:
\begin{itemize}
    \item $a_0$ is the Bohr radius.
    \item $r$ is the radial distance from the nucleus.
\end{itemize}

\textbf{Step 4: Angular Part $Y_{l,m}(\theta, \phi)$}

The angular part for the p orbitals (with $l = 1$) is described by the spherical harmonics $Y_{1,m}(\theta, \phi)$. The spherical harmonics for the p orbitals are:
\[
Y_{1,-1}(\theta, \phi) = \sqrt{\frac{3}{8\pi}} \sin(\theta) e^{-i\phi}
\]
\[
Y_{1,0}(\theta, \phi) = \sqrt{\frac{3}{4\pi}} \cos(\theta)
\]
\[
Y_{1,1}(\theta, \phi) = \sqrt{\frac{3}{8\pi}} \sin(\theta) e^{i\phi}
\]
\textbf{Step 5: Full Wavefunctions for $2p$ Orbitals}

Thus, the complete wavefunctions for the $2p$ orbitals (with $n = 2$, $l = 1$) are:

1. {For $m = -1$:}   
\[
   \psi_{2,1,-1}(r, \theta, \phi) = R_{2,1}(r) Y_{1,-1}(\theta, \phi)
   \]   
\[
   2p_{-1}=\psi_{2,1,-1}(r, \theta, \phi) = \frac{1}{8 \sqrt{\pi} }\left(\frac{1}{a_0}\right)^{5 / 2} r e^{-r / 2 a_0} \sin \theta e^{-i \phi}
   \]
2. {For $m = 0$:}   
\[
   \psi_{2,1,0}(r, \theta, \phi) = R_{2,1}(r) Y_{1,0}(\theta, \phi)
   \]   
\[
   2p_0 = \psi_{2,1,0}(r, \theta, \phi) = \frac{1}{4 \sqrt{2 \pi} }\left(\frac{1}{a_0}\right)^{5 / 2} r e^{-r / 2 a_0} \cos \theta
   \]
3. {For $m = 1$:}   
\[
   \psi_{2,1,1}(r, \theta, \phi) = R_{2,1}(r) Y_{1,1}(\theta, \phi)
   \]   
\[
   2p_1 = \psi_{2,1,1}(r, \theta, \phi) = \frac{1}{8 \sqrt{\pi} }\left(\frac{1}{a_0}\right)^{5 / 2} r e^{-r / 2 a_0} \sin \theta e^{i \phi}
   \]

\end{quote}

\subsection{Visualize the $2p$ orbitals }

We then proceed to request ChatGPT to provide a Python script for visualizing the orbitals. Due to the necessity of imposing multiple constraints on the specifics of the plotting process, we have decided to omit the detailed conversation with ChatGPT. The Python script for the orbital of $2p_{-1}$ generated by ChatGPT is presented in Listing \ref{listing1}.


\lstset{
    language=[LaTeX]TeX,
    comment=[l]{\#},
    commentstyle=\color{gray},
    basicstyle=\ttfamily,
    breaklines=true,
    showstringspaces=false,
}

\begin{lstlisting}[language=Python, caption=Python script to plot the probability density of the 2p\(_{-1}\) orbital, label=code:orbital, frame=single, basicstyle=\ttfamily\footnotesize, keywordstyle=\color{blue}, commentstyle=\color{gray}, label={listing1}]

import numpy as np
import matplotlib.pyplot as plt
from mpl_toolkits.mplot3d import Axes3D
from skimage import measure

# Constants
Z = 1  # Nuclear charge, for hydrogen Z = 1
a = 1  # Bohr radius, set to 1 for simplicity

# Define the probability density function in spherical coordinates
def psi_squared(r, theta, phi):
    psi = (1 / (8 * np.pi**0.5)) * (Z / a)**(5/2) * r * np.exp(-Z * r / (2 * a)) * np.sin(theta) * np.exp(-1j * phi)
    return np.abs(psi)**2

# Create a 3D grid in Cartesian coordinates
x = np.linspace(-10, 10, 100)
y = np.linspace(-10, 10, 100)
z = np.linspace(-10, 10, 100)
X, Y, Z_cart = np.meshgrid(x, y, z)

# Compute the spherical coordinates
R = np.sqrt(X**2 + Y**2 + Z_cart**2)
Theta = np.arccos(np.clip(Z_cart / R, -1, 1))  # Clip values to avoid invalid values due to numerical errors
Phi = np.arctan2(Y, X)

# Evaluate the probability density function on the grid
F = psi_squared(R, Theta, Phi)

# Find the isosurface for the probability density function
verts, faces, _, _ = measure.marching_cubes(F, level=0.0001)

# Convert vertices from the index space to actual coordinates
scale_factor = 20 / 99  # Conversion factor from index to actual coordinates
verts = verts * scale_factor - 10

# Plot the isosurface for the probability density function
fig = plt.figure(figsize=(10, 7))
ax = fig.add_subplot(111, projection='3d')

# Remove lines by setting edgecolor to 'none'
ax.plot_trisurf(verts[:, 0], verts[:, 1], verts[:, 2], triangles=faces, cmap='viridis', edgecolor='none', alpha=1)

# Set equal aspect ratio for all axes
ax.set_box_aspect([1, 1, 1])

# Set the limits for x, y, and z axes
ax.set_xlim([-10, 10])
ax.set_ylim([-10, 10])
ax.set_zlim([-10, 10])

ax.set_xlabel('x')
ax.set_ylabel('y')
ax.set_zlabel('z')
ax.set_title(r'Probability Density of $2p_{-1}$ Orbital')

# Save the plot as a PDF
plt.savefig('probability_density_plot.pdf', format='pdf')

plt.show()
\end{lstlisting}

\begin{figure*}[th]
    \centering
    \begin{subfigure}{0.3\textwidth}
        \centering
        \includegraphics[width=\textwidth]{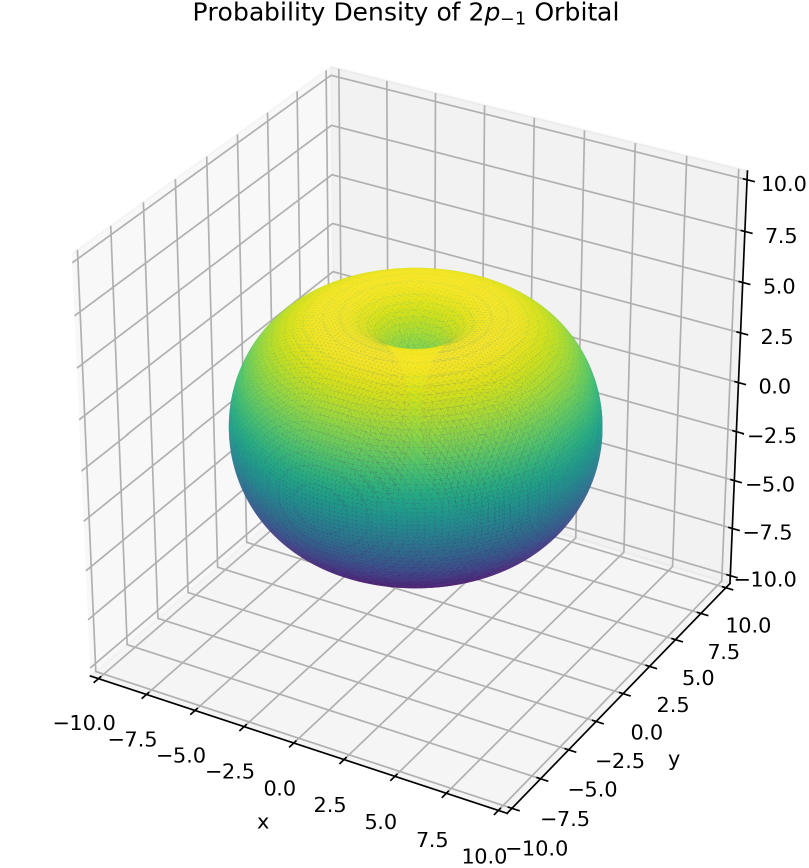}
        \label{fig:subfig1}
    \end{subfigure}
    \hfill
    \begin{subfigure}{0.3\textwidth}
        \centering
        \includegraphics[width=\textwidth]{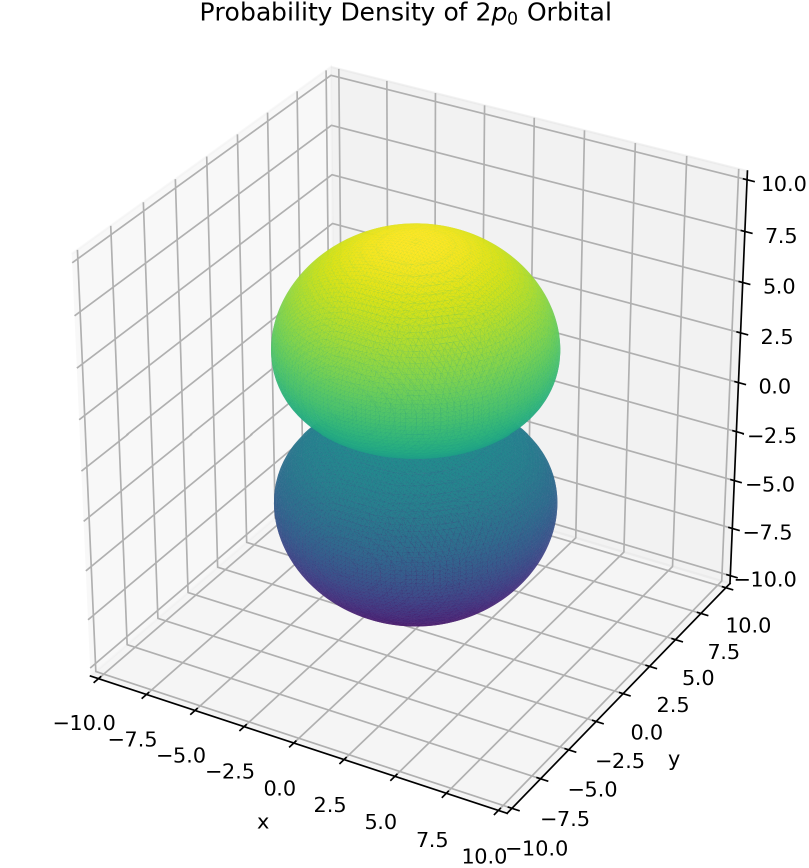}
        \label{fig:subfig2}
    \end{subfigure}
    \hfill
    \begin{subfigure}{0.3\textwidth}
        \centering
        \includegraphics[width=\textwidth]{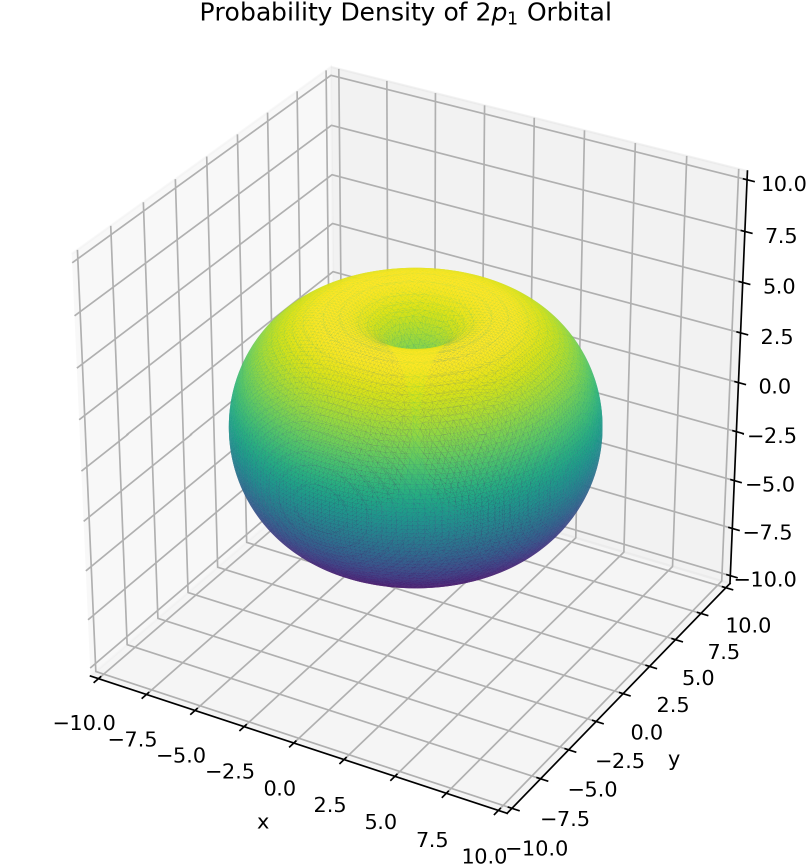}
        \label{fig:subfig3}
    \end{subfigure}

    \caption{The probability density of $2p_{-1}$, $2p_{0}$, and $2p_{1}$ orbitals.}
    \label{F1}
\end{figure*}

Subsequently, the script is executed within a Python Integrated Development Environment (IDE). For readers unfamiliar with the process of installing a Python IDE, ChatGPT can also provide a step-by-step installation guide. The remaining two $2p$ orbitals, namely $2p_{0}$ and $2p_{1}$, can be similarly visualized using the same script by substituting the corresponding wave functions, thereby generating a 3D visualization of these $2p$ orbitals, see Fig. \ref{F1}.

\begin{figure*}[th]
    \centering
    \begin{subfigure}{0.3\textwidth}
        \centering
        \includegraphics[width=\textwidth]{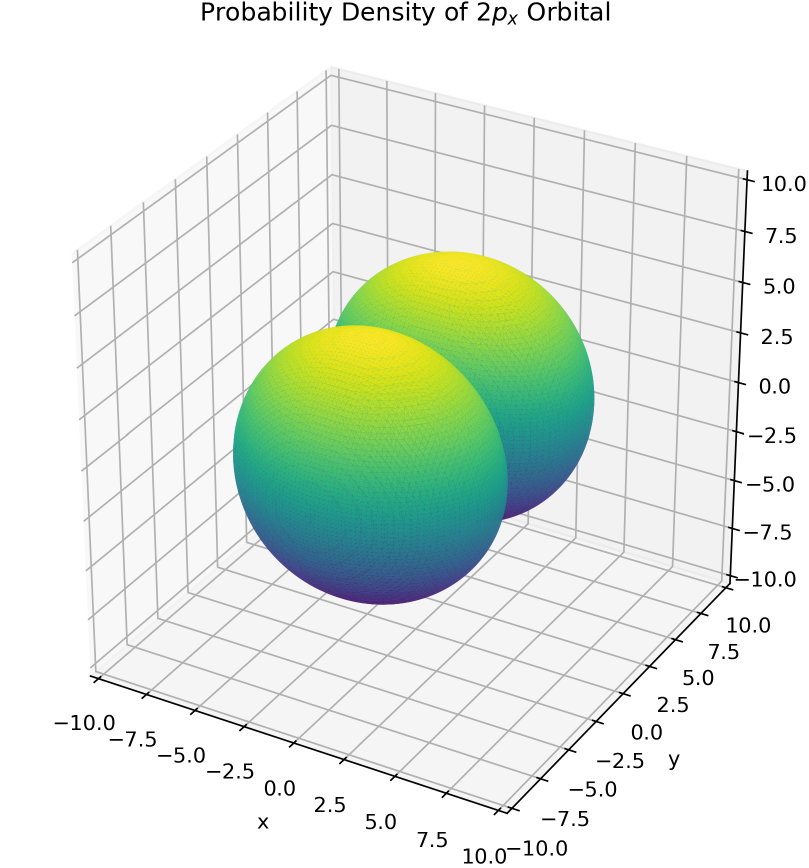}
        \label{fig2:subfig1}
    \end{subfigure}
    \hfill
    \begin{subfigure}{0.3\textwidth}
        \centering
        \includegraphics[width=\textwidth]{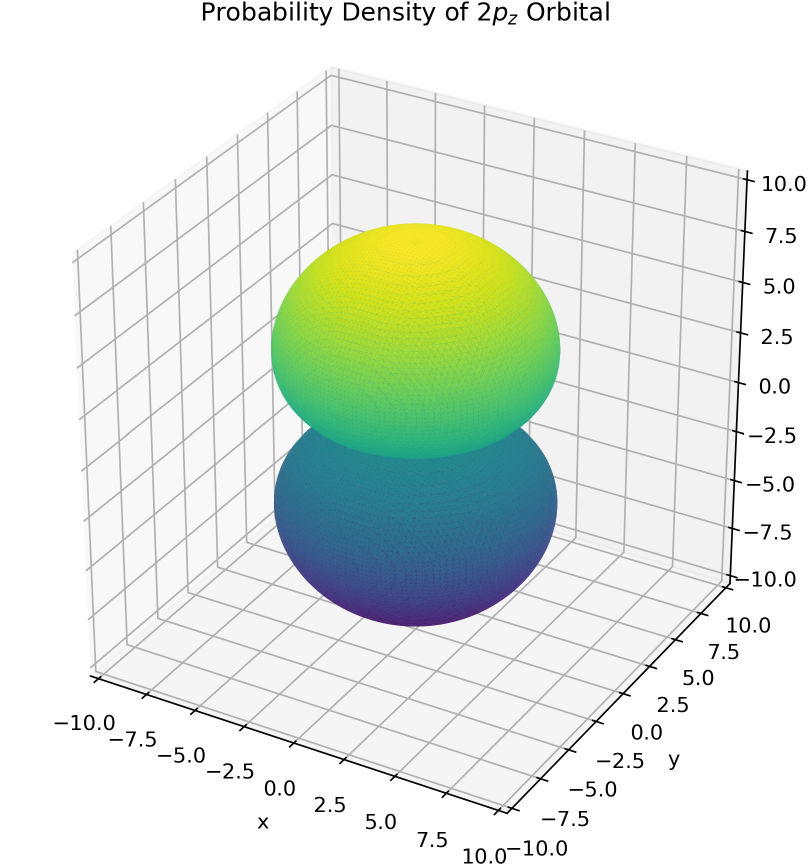}
        \label{fig2:subfig2}
    \end{subfigure}
    \hfill
    \begin{subfigure}{0.3\textwidth}
        \centering
        \includegraphics[width=\textwidth]{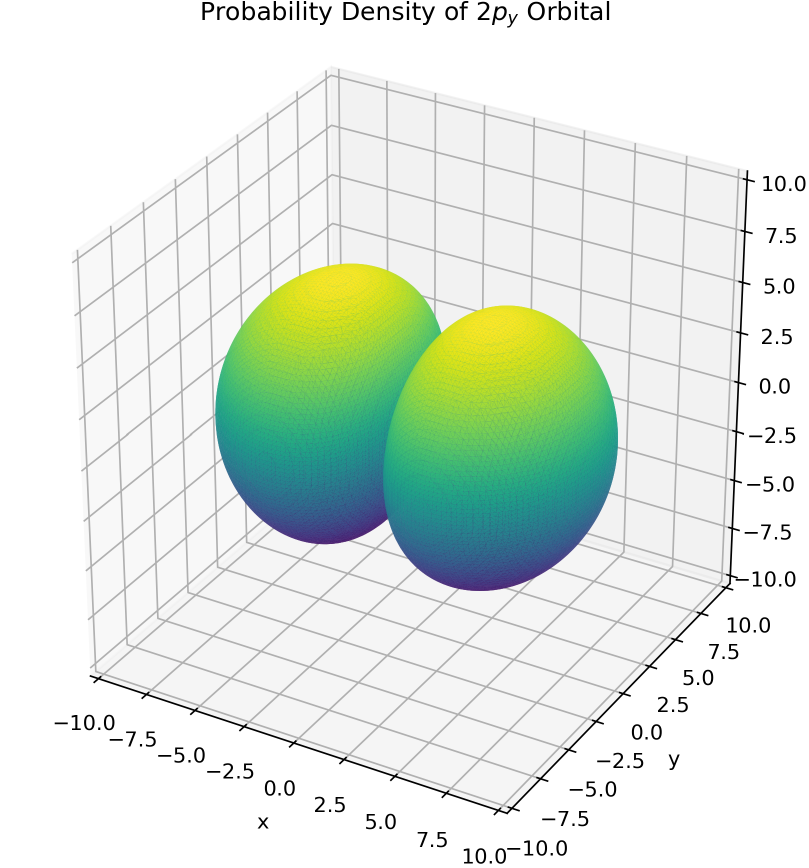}
        \label{fig2:subfig3}
    \end{subfigure}

    \caption{The probability density of $2p_{x}$, $2p_{y}$, and $2p_{z}$ orbitals.}
    \label{F2}
\end{figure*}

The three $2p$ orbitals presented here are evidently distinct from the commonly referenced $2p$ orbitals found in most standard textbooks (namely $2p_x$, $2p_y$, and $2p_z$). A closer analysis of the corresponding wave functions shows that $2p_{-1}$ and $2p_{1}$ possess identical real components (amplitudes) but differ in their phases (imaginary components). As a result, both orbitals share the same probability density, as depicted in Fig. \ref{F1}(a) and \ref{F1}(c). Due to the presence of complex terms in the wave functions, it is customary in the scientific community to express these orbitals as linear combinations. Any linear combination of these wave functions forms an eigenfunction of the Hamiltonian with the same energy eigenvalue. A conventional method to derive a real function from these wave functions is to combine them as follows:
\begin{equation}
\begin{aligned}
2 p_x &\equiv \frac{1}{\sqrt{2}}\left(2 p_{-1}+2 p_1\right) = \frac{1}{4 \sqrt{2 \pi}}\left(\frac{1}{a_0}\right)^{5 / 2} r e^{- r / 2 a_0} \sin \theta \cos \phi, \\
2 p_y &\equiv \frac{1}{i \sqrt{2}}\left(2 p_1-2 p_{-1}\right) = \frac{1}{4 \sqrt{2 \pi}}\left(\frac{1}{a_0}\right)^{5 / 2} r e^{- r / 2 a_0} \sin \theta \sin \phi , \\
2p_z &= 2p_0 = \frac{1}{4 \sqrt{2 \pi} }\left(\frac{1}{a_0}\right)^{5 / 2} r e^{-r / 2 a_0} \cos \theta.
\end{aligned}
\label{eq1}
\end{equation}
Subsequently, we visualize the $2p_x$, $2p_y$, and $2p_z$ orbitals by utilizing the same computational script, with the corresponding wave functions substituted accordingly. The resulting visualizations are presented in Fig. \ref{F2}. It is worth noting that while many textbooks also include similar 3D illustrations, generating them through a custom Python script offers the significant advantage of allowing these illustrations observable from all perspectives, thereby enhancing comprehension. Also note that, $2p_{-1}$ and $2p_{1}$ are eigenfunctions of $\hat{L}_z$ (with different eigenvalues: $-\hbar$ and $+\hbar$), while $2p_{x}$, $2p_{y}$ are not. This naturally leads to the next subsection, especially the principle of superposition.

\subsection{Symmetry, mixed state, and superposition}

The core concepts outlined above will be elucidated through an analysis of the questions posed in the abstract. Before addressing them, we first analyze the problem through the analogy of coin flipping. When a coin is tossed into the air, is it in which state: heads or tails? One could argue that, at that moment, the coin is in a superposition of both heads and tails. It is only upon landing (the act of measurement) that it collapses into a definite state, either heads or tails. The measurement thus destroys the superposition, causing the system to collapse into one particular state. This reflects the essence of a superposition: a superposition refers to a quantum system being in a linear combination of multiple basis states simultaneously. It applies to pure quantum states where the system exists in a coherent combination of several possible states. The ``axis" (here we could say that the heads and tails are directed to $+z$ and $-z$, respectively) is also defined by the measurement.

Next, we analyze the symmetry and mixed state. In the case of an ideal coin, heads and tails are symmetric. However, when one observes just one single coin after it has landed, it shows either heads or tails, seemingly breaking the heads-tails symmetry. Now, let us observe a large ensemble of coin tosses; the occurrences of heads and tails will each be approximately 50\%. This touches upon the concept of a mixed state. A mixed state represents a statistical ensemble of different quantum states, where the system is not in a definite quantum state but rather a probabilistic mixture of several possible states. Once the coin has landed, the statistical ensemble of outcomes (heads and tails) constitutes a typical example of a mixed state. Thus, symmetry is preserved in a statistical way, reflecting the true nature of symmetry. Symmetry, in this case, only holds at the level of mixed states.

Now, we jump into the case of $2p$ orbitals of a hydrogen atom. Based on the analogy mentioned above, the definite state of a $2p$ orbital must not possess spherical symmetry, akin to the head or tail states of a coin. However, the mixed state of all the $2p$ orbitals in equal proportions should be spherically symmetric. 

Again, we can request ChatGPT to prove this mathematically, the conversation is as follows:

\begin{quote}
\textbf{User:} Please prove that the mixed state of $2p$ orbitals in equal proportions is spherically symmetric. \\
\textbf{ChatGPT:} We are considering a mixed state of the three \(2p\) orbitals, meaning the system has an equal probability of being in one of the \(2p_x\), \(2p_y\), or \(2p_z\) orbitals. In a mixed state, the probability densities are treated separately, not as a superposition of the wavefunctions.

\textbf{Step 1: Probability Density for Each \(2p\) Orbital}

The angular part of the wavefunctions for the three \(2p\) orbitals are:
\(-2p_x\) orbital: \(\psi_{2p_x} \propto r \sin \theta \cos \phi\)  
\[
  |\psi_{2p_x}|^2 \propto r^2 \sin^2 \theta \cos^2 \phi
  \]
 \(-2p_y\) orbital: \(\psi_{2p_y} \propto r \sin \theta \sin \phi\)  
\[
  |\psi_{2p_y}|^2 \propto r^2 \sin^2 \theta \sin^2 \phi
  \]
 \(-2p_z\) orbital: \(\psi_{2p_z} \propto r \cos \theta\)  
\[
  |\psi_{2p_z}|^2 \propto r^2 \cos^2 \theta
  \]
\textbf{Step 2: Mixed State Probability Distribution}

In a mixed state, the total probability density \(P_{\text{mixed}}\) is the weighted sum of the probability densities for each orbital. Since we have equal proportions of the \(2p_x\), \(2p_y\), and \(2p_z\) orbitals, the mixed state's total probability density is:
\[
P_{\text{mixed}} = \frac{1}{3} \left( |\psi_{2p_x}|^2 + |\psi_{2p_y}|^2 + |\psi_{2p_z}|^2 \right)
\]
Substituting the probability densities for each orbital:
\[
P_{\text{mixed}} \propto \frac{1}{3} \left( r^2 \sin^2 \theta \cos^2 \phi + r^2 \sin^2 \theta \sin^2 \phi + r^2 \cos^2 \theta \right)
\]
\textbf{Step 3: Simplifying the Expression}

Factor out \(r^2\) from the equation:
\[
P_{\text{mixed}} \propto r^2 \left( \frac{1}{3} \left( \sin^2 \theta (\cos^2 \phi + \sin^2 \phi) + \cos^2 \theta \right) \right)
\]
Using the trigonometric identity \(\cos^2 \phi + \sin^2 \phi = 1\), this simplifies to:
\[
P_{\text{mixed}} \propto r^2 \left( \frac{1}{3} \left( \sin^2 \theta + \cos^2 \theta \right) \right)
\]
Since \(\sin^2 \theta + \cos^2 \theta = 1\), we get:
\[
P_{\text{mixed}} \propto r^2 \times \frac{1}{3} \times 1 = \frac{r^2}{3}
\]
\textbf{Step 4: Interpretation}

The mixed state's probability density depends only on \(r^2\), the radial distance from the nucleus, and is independent of the angles \(\theta\) and \(\phi\). This implies that the electron density is spherically symmetric in the mixed state.
\end{quote}

\begin{figure}[thb]
\centering
	\includegraphics[width=\linewidth]{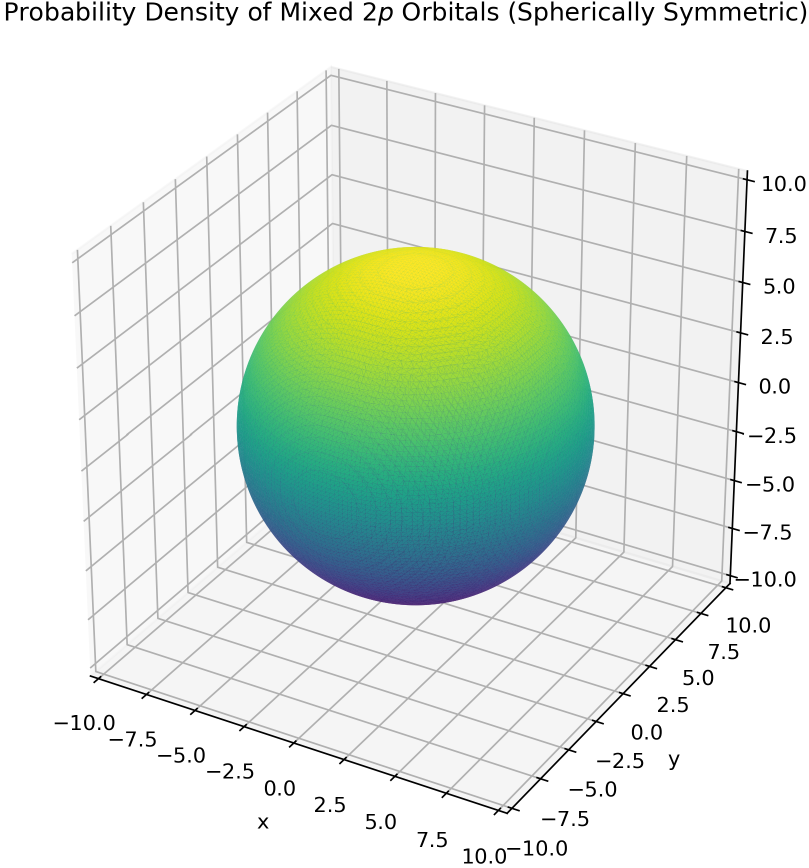}
	\caption{The spherically symmetric probability density of $2p$ orbitals in the mixed state of equal proportion.}
	\label{F3}
\end{figure}

Now we applied the ChatGPT-generated Python script again, by replacing the wave function with the flowing mixed state:
\begin{equation}
    P_{\text{mixed}} = \frac{1}{3} \left( |\psi_{2p_x}|^2 + |\psi_{2p_y}|^2 + |\psi_{2p_z}|^2 \right)
\end{equation}
this directly gives a spherically distributed probability density, see Fig. \ref{F3}

\bigskip

Finally, we can address the questions posed in the abstract. The $p$ orbitals exhibit a dumbbell shape due to the angular momentum quantum number $l=1$, which results in a wave function dependent on the directional angles. For a hydrogen atom with an electron in the $2p$ state, the specific $2p$ orbital occupied by the electron is not uniquely defined unless additional information is provided, such as the spatial orientation of the orbital or the presence of an external field. This is because the three $2p$ orbitals ($2p_x$, $2p_y$, and $2p_z$) are degenerate. Therefore, the electron can occupy any linear combination (superposition) of these orbitals. The axes associated with the $2p_x$, $2p_y$, and $2p_z$ orbitals are determined by the measurements (or how we prepare the $2p$ state, which can also be regarded as a measurement). In the absence of measurement, the $x$, $y$, and $z$ axes lack any intrinsic relationship to the real world. When we measure the probability density of such a hydrogen atom, the resulting distribution will typically exhibit a dumbbell shape. However, the weighted average of all possible $2p$ orbitals (if the $2p$ states were randomly stimulated) yields a spherically symmetric distribution. Similarly, this also applies to $d$ and $f$ orbitals.

\section{Conclusion}

In conclusion, AI tools such as ChatGPT can play a pivotal role in helping students grasp complex quantum mechanics concepts that often pose significant challenges. This paper explores the common confusion regarding the shape of $p$ orbitals, and how it relates to the underlying principles of symmetry, mixed state, and superposition in quantum mechanics. This approach has the potential to significantly enhance the learning experience for undergraduate students, providing valuable insights into quantum mechanics and supporting educators in making difficult concepts more accessible.

\appendix

\section*{APPENDIX: The critical role of measurement}
A specific quantum state of a system can be manipulated through measurement. For $2p$ orbitals of a hydrogen atom, if we measure the orbital shape along the $x$ axis (defined by the orientation of the measuring apparatus), then we can collapse the state into $2p_x$. Subsequently, using a Stern-Gerlach apparatus (SGA) on the $z$ axis, we obtain $2p_{-1}$ and $2p_{1}$ with equal probability (the electrons will split along the $z$ axis). This can be understood by recalling the following relationship from Eq. \ref{eq1}, $2 p_x = \frac{1}{\sqrt{2}}\left(2 p_{-1}+2 p_1\right)$. A simpler example of state manipulation can be seen in spin-$1/2$ systems, where multiple SGAs can be used to control the spin state in a similar manner, which is considered as the foundation for quantum computing \cite{hughes2021}.


\begin{thebibliography}{3}%
\makeatletter
\providecommand \@ifxundefined [1]{%
 \@ifx{#1\undefined}
}%
\providecommand \@ifnum [1]{%
 \ifnum #1\expandafter \@firstoftwo
 \else \expandafter \@secondoftwo
 \fi
}%
\providecommand \@ifx [1]{%
 \ifx #1\expandafter \@firstoftwo
 \else \expandafter \@secondoftwo
 \fi
}%
\providecommand \natexlab [1]{#1}%
\providecommand \enquote  [1]{``#1''}%
\providecommand \bibnamefont  [1]{#1}%
\providecommand \bibfnamefont [1]{#1}%
\providecommand \citenamefont [1]{#1}%
\providecommand \href@noop [0]{\@secondoftwo}%
\providecommand \href [0]{\begingroup \@sanitize@url \@href}%
\providecommand \@href[1]{\@@startlink{#1}\@@href}%
\providecommand \@@href[1]{\endgroup#1\@@endlink}%
\providecommand \@sanitize@url [0]{\catcode `\\12\catcode `\$12\catcode `\&12\catcode `\#12\catcode `\^12\catcode `\_12\catcode `\%12\relax}%
\providecommand \@@startlink[1]{}%
\providecommand \@@endlink[0]{}%
\providecommand \url  [0]{\begingroup\@sanitize@url \@url }%
\providecommand \@url [1]{\endgroup\@href {#1}{\urlprefix }}%
\providecommand \urlprefix  [0]{URL }%
\providecommand \Eprint [0]{\href }%
\providecommand \doibase [0]{https://doi.org/}%
\providecommand \selectlanguage [0]{\@gobble}%
\providecommand \bibinfo  [0]{\@secondoftwo}%
\providecommand \bibfield  [0]{\@secondoftwo}%
\providecommand \translation [1]{[#1]}%
\providecommand \BibitemOpen [0]{}%
\providecommand \bibitemStop [0]{}%
\providecommand \bibitemNoStop [0]{.\EOS\space}%
\providecommand \EOS [0]{\spacefactor3000\relax}%
\providecommand \BibitemShut  [1]{\csname bibitem#1\endcsname}%
\let\auto@bib@innerbib\@empty
\bibitem [{\citenamefont {Henderson}(2023)}]{PhysRevPhysEducRes.19.020003}%
  \BibitemOpen
  \bibfield  {author} {\bibinfo {author} {\bibfnamefont {C.}~\bibnamefont {Henderson}},\ }\href {https://doi.org/10.1103/PhysRevPhysEducRes.19.020003} {\bibfield  {journal} {\bibinfo  {journal} {Phys. Rev. Phys. Educ. Res.}\ }\textbf {\bibinfo {volume} {19}},\ \bibinfo {pages} {020003} (\bibinfo {year} {2023})}\BibitemShut {NoStop}%
\bibitem [{\citenamefont {Levine}(2014)}]{levine2014}%
  \BibitemOpen
  \bibfield  {author} {\bibinfo {author} {\bibfnamefont {I.~N.}\ \bibnamefont {Levine}},\ }\href@noop {} {\emph {\bibinfo {title} {Quantum Chemistry}}}\ (\bibinfo  {publisher} {Pearson Prentice Hall Upper Saddle River, NJ},\ \bibinfo {year} {2014})\BibitemShut {NoStop}%
\bibitem [{\citenamefont {Hughes}\ \emph {et~al.}(2021)\citenamefont {Hughes}, \citenamefont {Isaacson}, \citenamefont {Perry}, \citenamefont {Sun},\ and\ \citenamefont {Turner}}]{hughes2021}%
  \BibitemOpen
  \bibfield  {author} {\bibinfo {author} {\bibfnamefont {C.}~\bibnamefont {Hughes}}, \bibinfo {author} {\bibfnamefont {J.}~\bibnamefont {Isaacson}}, \bibinfo {author} {\bibfnamefont {A.}~\bibnamefont {Perry}}, \bibinfo {author} {\bibfnamefont {R.~F.}\ \bibnamefont {Sun}},\ and\ \bibinfo {author} {\bibfnamefont {J.}~\bibnamefont {Turner}},\ }\href@noop {} {\emph {\bibinfo {title} {Quantum Computing for the Quantum Curious}}}\ (\bibinfo  {publisher} {Springer Nature},\ \bibinfo {year} {2021})\BibitemShut {NoStop}%
\end{thebibliography}
\end{document}